\documentclass{IEEEtran}
\usepackage[utf8]{inputenc}
\usepackage{graphicx,
	psfrag,
	epsfig,
	epstopdf,
	amsthm,
	amssymb,
	url,
	subcaption,
	algorithm,
	algorithmic,
	balance,
	enumerate,
	color,
    url,
	setspace,
	tikz,
	pgfplots
}
\usepackage{amsmath}%[centertags][tbtags][sumlimits][nosumlimits][intlimits][namelimits][nonamelimits]
\usepackage[nospace,noadjust]{cite}
\usepackage{pbox}
\usepackage{geometry}
\usepackage{multirow}
\usepackage{adjustbox}
\usepackage{array}
\usepackage{booktabs}
\usepackage{float}

\newcommand{\cref}[1]{Constraint~\ref{#1}}

\newcommand{\ignore}[1]{}
\newgeometry{left=1.2cm,right=1.2cm,bottom=1.5cm,top=1.5cm}

\begin{document}

\title{IoT-based Wearables: A comprehensive Survey}

	\author{
	\IEEEauthorblockN{Yahuza Bello \IEEEauthorrefmark{1}, Emanuel Figetakis \IEEEauthorrefmark{1}}

	\IEEEauthorblockA{\IEEEauthorrefmark{1} University of Guelph, Ontario, Canada.}}

\maketitle
\begin{abstract}
A substantial amount of growth is being achieved by businesses through IoT-based services. The emergent of small electronic devices capable of computing, which are commonly known as wearables in IoT domain has proven to have huge impact in people's life. Theses wearables are capable of collecting vital information about a person's activities and behaviours regularly. This makes them suitable for many applications in health monitoring, fitness, sports, education and some industry related applications. To this end, in this paper, we aim to provide a general review on IoT-based wearables, the sensors adopted for several categorized wearables, the communication technologies adopted and the most widely adopted data processing techniques for wearables. Furthermore, we present the challenges faced for wide adoption of wearables and the future research directions.
\end{abstract}

\begin{IEEEkeywords}
Internet of things, wearables, communication technologies, data analytic
\end{IEEEkeywords}

\section{Introduction}
\begin{figure*}[h]
    \centering
    \includegraphics[width=1\textwidth]{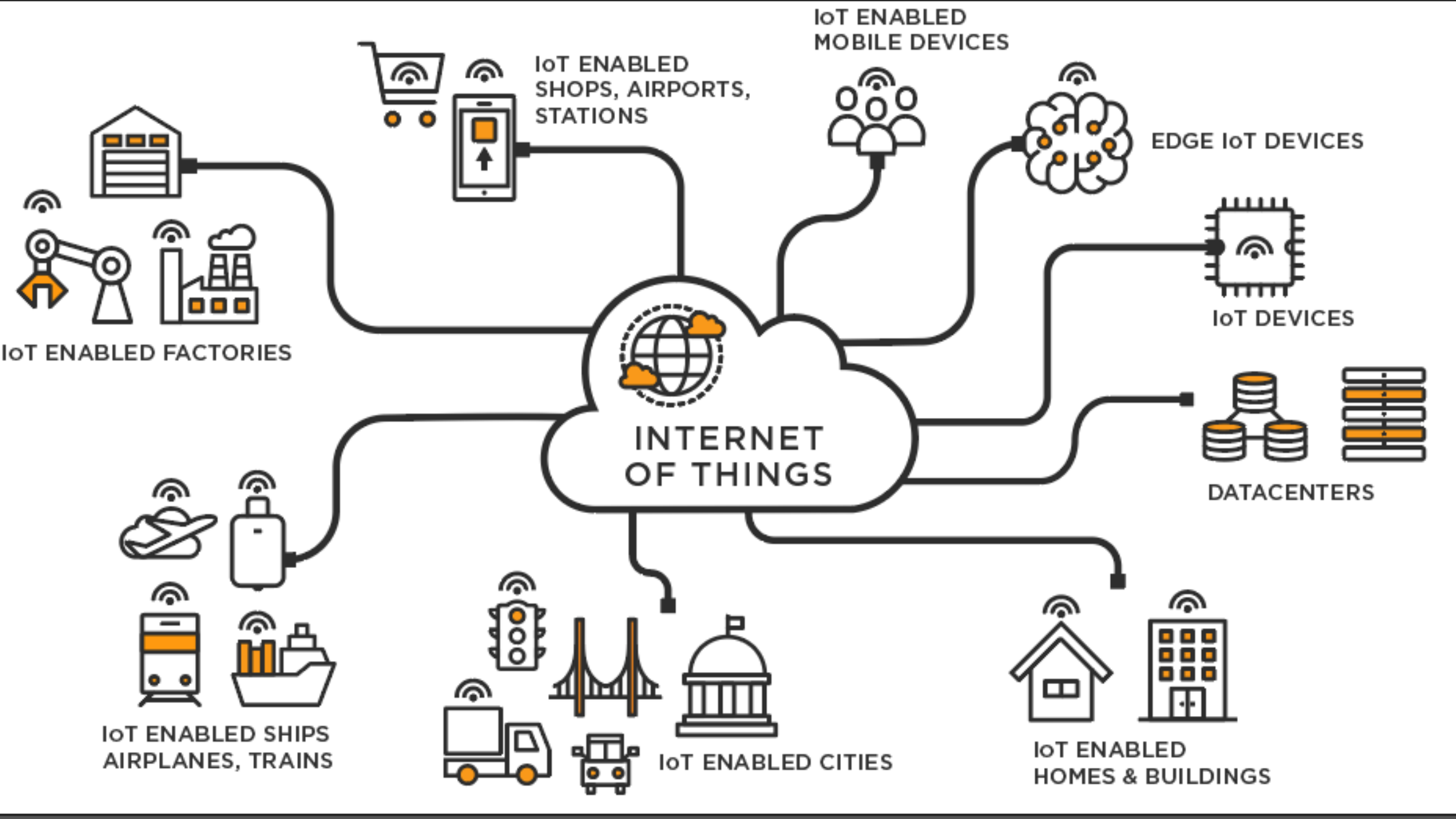}
    \caption{Internet of things Applications \cite{iotpic}}
    \label{fig:iotapp}
\end{figure*}

\IEEEPARstart{I}{n} today's modern era, several applications are developed with the need of connectivity to the internet to function effectively. The Internet of Things (IoT) is the concept of having embedded systems interconnected together and communicating through the Internet \cite{6714496}. This provides the required connectivity to the internet for such applications. The IoT domain has been thoroughly investigated in both academia and industry and its development through the years was rapid because of its wide range of applications in various fields such as health, IoT-enabled sports, IoT-enabled factories, agriculture, IoT-enabled cities, traffic management, smart supply chain and smart grid to name a few as depicted in Figure~\ref{fig:iotapp}. 

\begin{figure}[h]
    \centering
    \includegraphics[width=.5\textwidth]{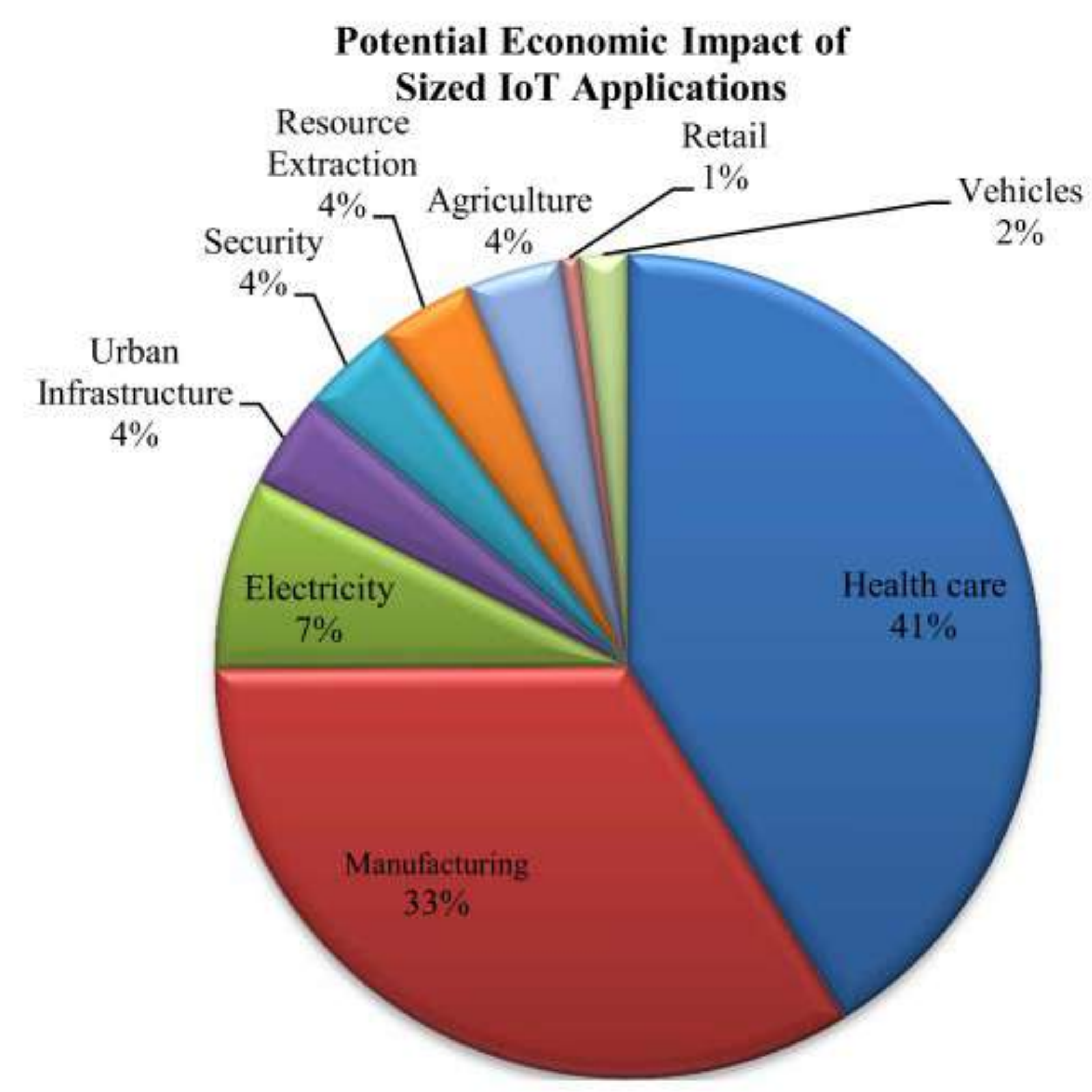}
    \caption{Projected market share of dominant IoT applications by 2025 \cite{7123563}}
    \label{fig:iotmarkt}
\end{figure}

A substantial amount of growth is being achieved by businesses through IoT-based services. For example, the biggest economic impact will be from healthcare and manufacturing applications. The global economy is expected to generate about $1.1-$2.5 trillion US dollars in growth annually by 2025 as a result of IoT-based healthcare applications and services, such as mobile health (mHealth) and telecare services that improves diagnosis, treatment and other monitoring services. IoT is estimated to generate a total economic impact amounting to \$2.7 trillion - \$6.2 trillion annually by 2025 \cite{chui2015four}. In Figure~\ref{fig:iotmarkt}, dominant IoT applications (i.e., applications in healthcare, manufacturing, electricity, urban infrastructure, security, resource extraction, agriculture, retail and vehicles) are shown with their projected market share \cite{7123563}.

\begin{figure*}[h]
    \centering
    \includegraphics[width=1\textwidth]{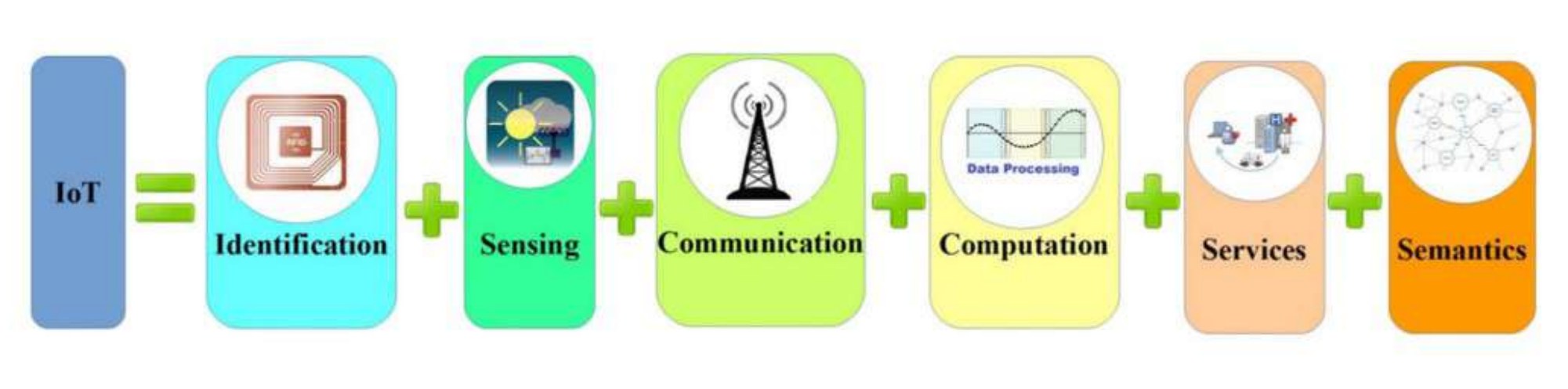}
    \caption{IoT components \cite{7123563}}
    \label{fig:iotelements}
\end{figure*}

A typical IoT consists of 6 building blocks as shown in Figure~\ref{fig:iotelements}. These buildings blocks are identification, sensing, communication, computation/processing, services and semantics. By knowing these building blocks, you can gain a deeper understanding of how IoT works and what it means. Identifying services and matching them with customer demands is crucial for IoT. Electronic product codes (EPCs) and ubiquitous codes (uCodes) are two most adopted identification methods for the IoT systems. Sensors in the IoT gather data from related objects and send it to data warehouses, databases, or the cloud. Based on the analysis performed on the collected data, specific actions are taken according to the services that are required. There are a variety of IoT sensors, including smart sensors, actuators, and wearable sensors. We will discuss sensors in the context of wearable devices in section IV. Using multiple communication technologies, heterogeneous IoT objects can be connected to provide specific smart services. The different communication technologies used in the context of wearable devices will be discussed in detail in section V. In IoT, processors, microcontrollers, System-on-Chips (SoCs), and Field-Programmable Gate Arrays (FPGAs) comprise the "brain" and computational power. Arduino, UDOO, FriendlyARM, Intel Galileo, Raspberry PI, Gadgeteer, BeagleBone, Cubieboard, Z1, WiSense, Mulle, and T-Mote Sky are some of the hardware platforms designed to run IoT applications.

The emergent of small electronic devices capable of computing, which are commonly known as wearables in IoT domain has proven to have huge impact in peoples life \cite{NIKNEJAD2020103529}. Theses wearables are capable of collecting vital information about a person's activities and behaviours regularly. This makes them suitable for many applications in health monitoring, fitness, sports, education and some industry related applications. Wearables are often worn as additional accessories on individual's clothing, as implants on certain part of the body, or even tattooed to the skin. Being part of the IoT ecosystem, wearables are connected to the Internet in order to gather and transmit vital information. Additionally, the mobility of people and animals makes wearable devices increasingly important since they can collect, send, and receive data from the Internet in real time, and thus help us to make better decisions to improve our lifestyle.

\begin{figure}[h]
    \centering
    \includegraphics[width=.5\textwidth]{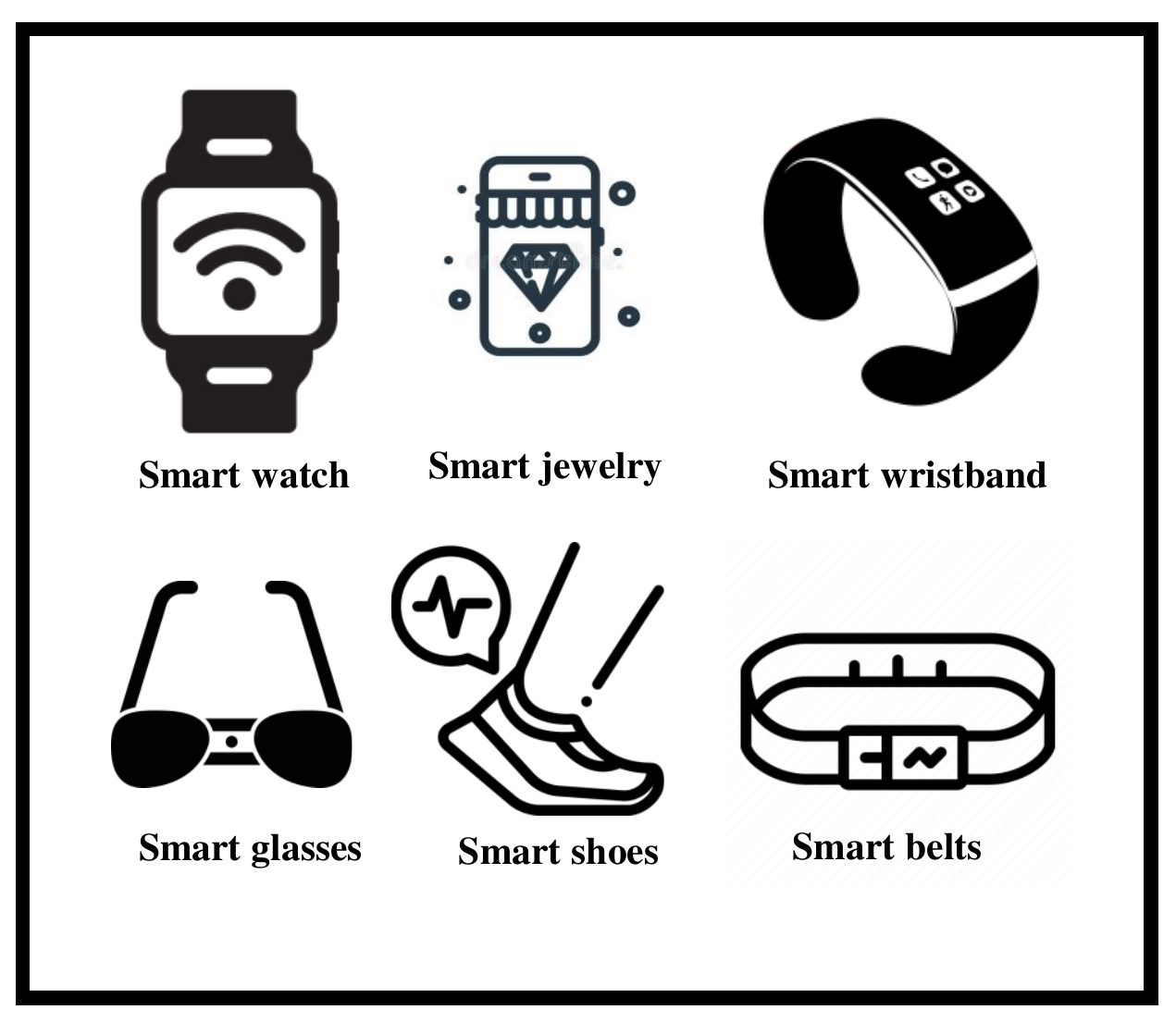}
    \caption{IoT-based Wearable products}
    \label{fig:wicons}
\end{figure}

A wide variety of wearable products such as smart jewelleries, smart wristbands, smart watches, smart glasses, smart shoes and smart belts are already available in the markets as illustrated in Figure~\ref{fig:wicons}. According to International Data Corporations (IDC), the global shipments of wearable products are projected to grow from 66.5 million units as reported in 2019 to 105.3 million units by the end of 2024 \cite{idc}. Wearable devices first emerged as fitness activity trackers, then comes other applications such as Bluetooth headsets, smartwatches, and web-enabled glasses shortly \cite{NAHAVANDI2022106541}. Afterwards, virtual reality headsets and augmented reality headsets were introduced in the gaming industry. However, health monitoring and medical usage cases are the most important life-changing applications of IoT-based wearable technologies.

Wearable devices are a rapidly developing field with the potential to open up new applications. This review is motivated by the potential of wearable devices to be able to open up new applications in various fields. Specifically, in this review, we will focus mainly on identifying the different wearables in the IoT domain, discuss the different sensors adopted for these wearables and then cover the different data processing techniques and communication technologies that are widely used for the IoT-based wearables. 

Although there are several survey papers \cite{NIKNEJAD2020103529,pourbemany2021survey,bios11100372,article1, 12,s21155218,s21165418} that addressed the concept of wearables and the ongoing research efforts, most of them focused on specific application such as applications in fitness and health monitoring neglecting the sensors behind each wearable, the various communication techniques adopted and data processing techniques that are commonly used for IoT-based wearables. To this end, in this paper, we aim to provide a general review on IoT-based wearables, the sensors adopted for several categorized wearables, the communication technologies adopted and the most widely adopted data processing techniques for wearables.

The rest of the paper is structured as follows: Section II presents the relevant survey papers published in the literature. Section III introduces IoT-based wearables and categorized them according to the targeted application. In section IV, the most commonly used sensors for IoT-based wearables are discussed. Section V discusses the various communication technologies adopted for wearables and the various data analytic techniques used for data processing. Section VI presents the challenges faced for wide adoption of wearables and the future research directions. Section VII concludes the paper.

\section{Related Works}

A number of other research studies have examined wearable technologies in different ways. For example, the survey paper in \cite{NIKNEJAD2020103529} provides academics seeking further research topics with an in-depth understanding of smart wearables concept. The authors present an analysis of behavioral predictors of smart wearable adoption and adoption intentions among people. In \cite{pourbemany2021survey}, using signals and sensors, the authors examines context-based pairing in wearable devices. The author in \cite{bios11100372} examine each type of IoT architecture as well as different methods of data transfer, data processing, and computing paradigms. Then the IoT-assisted wearable sensor systems and their various applications in healthcare as well as the various communication technologies adopted are presented in the paper. The authors in \cite{article1} reviews both scientific papers and commercial efforts related to wearable health care devices. The paper focuses on the most important wearable devices that directly measure health status parameters. An overview of existing literature on intelligent wearables is presented in \cite{12}. The authors went further to provide a review of the risks of using intelligent wearables and explain what risks were considered in previous research. The authors in \cite{s21155218} discuss and explore several communication and artificial intelligence techniques, which are suitable for the next generation of wearable devices. These techniques when fully adopted will enable emergence of innovative services. An overview of recent research on wearables and IoT technologies used for fitness assessment is provided in \cite{s21165418}.

\section{Wearables in IoT}

There are many applications in the field of IoT that can be enhanced by wearable technology. However, wearable devices will become worth their weight in gold when they are integrated into a true IoT system. As a result, most research papers currently published in the literature connect wearable devices to the Internet in one of two ways. Either the wearable devices send data to the cloud or to an Internet server for offline processing or some of the data processing is done locally on the wearable device. Having integrated IoT platforms and addressing many issues pertaining to data ownership, data sharing policies, privacy, and safety will fully realize the prevalence of wearable devices. 

In many cases, a mobile network is needed to support these wearable devices. Data is also a concern when dealing with sensors that collect many readings. In works \cite{ybello1,ybello2,ybello3,ybello4,ybello5} the authors create specialized networks that can handle the needs of wearable devices. These networks also address several concerns such as security and data storage. In, \cite{efigetakis} shows that edge nodes can also play an important role with IoT devices and data collection. It can reduce load on the server and save storage space. 

Several research works have categorized IoT-based wearables by considering multiple factors. As part of its ongoing effort to standardize wearable electronic devices and technologies, Technical Commission (TC) 124 of the IEC (International Electrotechnical Commission) identifies four types of smart wearables: accessory wearables, textile/fabric wearables, patchable wearables and implantable wearables. For details on this categorization, refer to \cite{idccc}. The IEC Standardization Group (SG) 10 on wearable smart devices also indicated that these wearable devices can be categorized in accordance with their location within, on, or near an individual. The categories are near-body wearables, in-body wearables, on-body wearables and electronic textiles \cite{iec}. According to \cite{seneviratne2017survey}, IoT-based wearable devices can be categorized into wrist-worn devices (such as smart watches and wrist bands), haed-mounted devices (such as smart eyewear, headset and earbuds), e-textile (such as smart garments), e-patches (such as sensor patches and e-tattoo) and others (this category covers that do not fall under the other categories such as smart jewellery and straps).

\begin{figure}[h]
    \centering
    \includegraphics[width=.5\textwidth,  height=6.5cm]{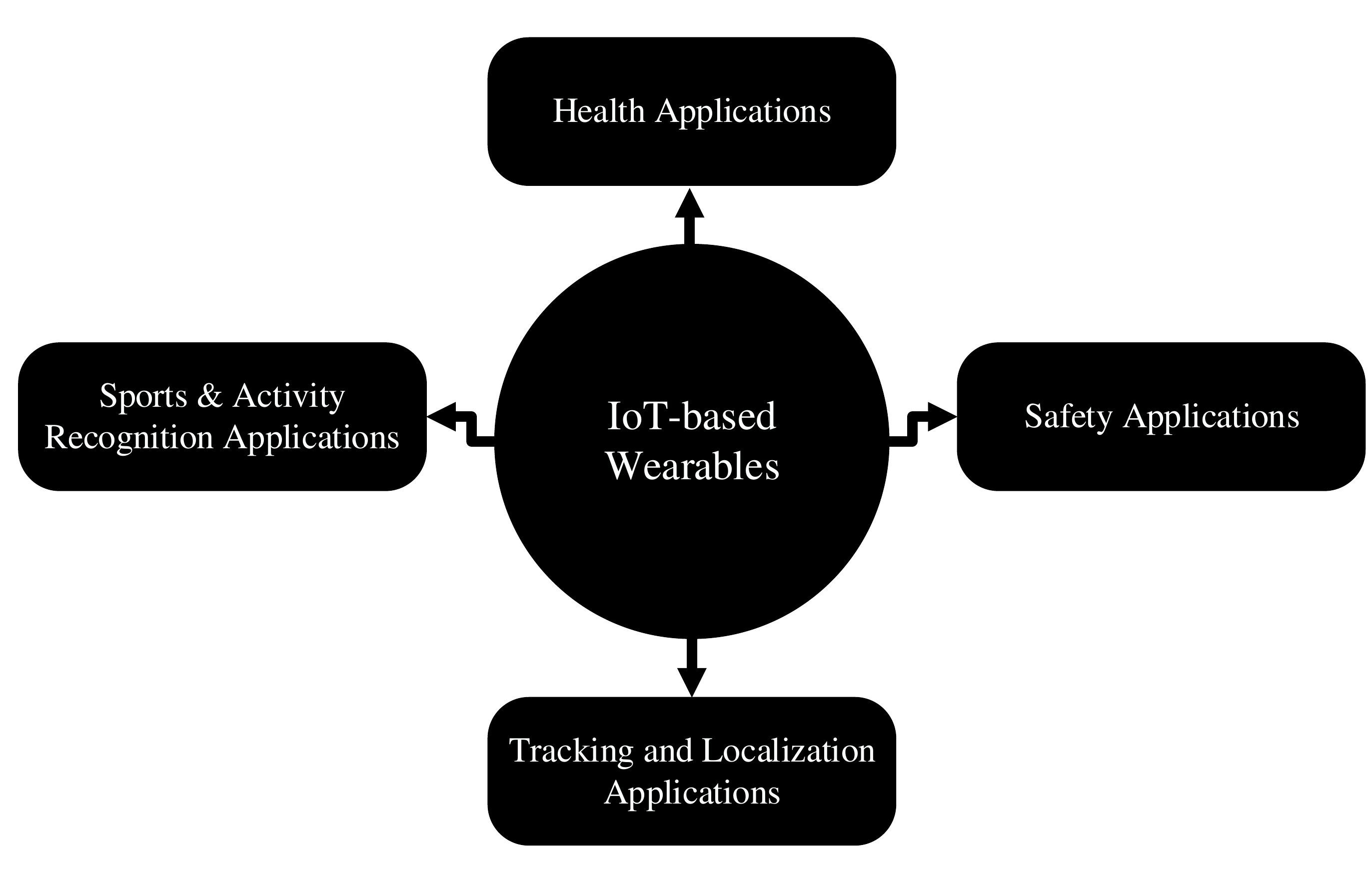}
    \caption{Application-based Classification of Wearable Devices}
    \label{fig:wcateg}
\end{figure}

In this paper, we categorized wearable devices according to the targeted applications in IoT domain similar to the work in \cite{dian2020wearables}. As depicted in Figure~\ref{fig:wcateg}, IoT-based wearbles can be classified according to the targeted applications as Health-based wearables, safety-based wearables, sports and activity recognition wearables and tracking and localization wearables.

\setlength{\arrayrulewidth}{0.5mm}
\setlength{\tabcolsep}{10pt}
\renewcommand{\arraystretch}{1.5}
\begin{table*}[ht!]
\begin{center}
\caption{Summary of the Categorized Wearables and the relavant related works}
\begin{tabular}{  |m{12em} | m{25em}| m{7em}|   }
  \hline
 Wearable categories & Description & Related works \\
  \hline
 Health-based applications & Wearable IoT devices used to monitor and treat patients remotely, and in some cases for rehabilitation. & \cite{dian2020wearables,nave2018smart,yang2018iot,xu2017design,xu2014ubiquitous,ghorbel2018cloud,agyeman2019design}  \\
  \hline
  Safety Applications & wearables in the safety category are used to provide a safe environment for users. & \cite{nyan2008wearable, min2018detection, pierleoni2015high, ozcan2016autonomous, ozcan2014automatic, dhole2019novel}  \\
  \hline
  Sports and activity recognition & This category pertains to applications where wearables are worn during sport activities to track different metrics of user/athlete activity to improve their performance. & \cite{reyes2016transition,qi2018hybrid}  \\
  \hline
 Tracking and localization & Tracking people and animals online is the most common use of this category. & \cite{shit2018location,tomic2014rss}  \\
  \hline
\end{tabular}
\label{table:SDPZTA}
\end{center}
\end{table*}

\subsection{Health-based wearables}
In the health sector, Wearable IoT devices are usually used to monitor and treat patients remotely, and in some cases for rehabilitation. Data about a patient's health is collected through sensors, and the wearable device may perform small analysis before sending the data to the Internet for further processing. Additionally, these wearables are capable of receiving additional inputs to aid for further analysis. Often, health-based wearable devices are typically connected to smartphones for analyzing data and sending it to cloud computing frameworks like Azure or Amazon Web Services (AWS) for handling, storing, and processing. The use of mobile health applications can provide insight about a patient's health and provide a visual representation of the analysis of the data. The wearable also can be programmed to respond to special commands, such as heating up the body or applying shocks, based on the analyzed data during treatment.

Health-based applications of wearables can be broadly categorized as health treatment and rehabilitation wearable systems and health monitoring wearable systems \cite{dian2020wearables}. Several studies have demonstrated the significance of health treatment and rehabilitation wearable systems. 

By using rehabilitation wearables, disabled patients can maintain or improve their mental or physical abilities. The authors in \cite{nave2018smart} presents a walker-based physiotherapy system for monitoring and evaluating movement metrics is proposed, which sends the data to the cloud for analysis, and displayed on a mobile application in real time. In \cite{yang2018iot}, a stroke rehabilitation framework that utilizes smart wearable armbands, machine learning algorithms, and 3-D-printed robot hands to assist stroke patients was proposed. An IoT sensing device was integrated with textile electrodes to develop the SWA, which measures, pre-processes, and wirelessly transmits biopotential information. The work in \cite{xu2017design} presents a framework for an m-Health monitoring system based on a cloud computing platform (Cloud-MHMS) for achieving personalised and high-quality health monitoring using new technologies, such as mobile networks and cloud computing. A system using IoT data to support emergency medical services is presented as a demonstration of how data can be collected, integrated, and interoperated flexibly in \cite{xu2014ubiquitous}. The authors develop a Resource-based Data Accessing-IoT (UDA-IoT) method that allow access to IoT data resources in a universal manner in order to enhance the accessibility of IoT data resources. The authors in \cite{ghorbel2018cloud} propose a new cloud-based wheelchair assist system, which supports impaired drivers by providing safe driving conditions. The authors utilized an embedded mobile System-On-a-Chip (SoC) and android based mobile application for the proposed system. A gaming-based rehabilitation system integrating wearable technology and IoT is presented in \cite{agyeman2019design} in order to help stroke patients who, suffer from upper limb disabilities.

In the same fashion, health monitoring wearable systems are categorized based on the kind of sensor adopted as Bio-potential sensors, motion sensors, environmental sensors and chemical sensors \cite{dian2020wearables}. The different kind of sensors used in most IoT-based wearables will be discussed in detail in the next section.

\setlength{\arrayrulewidth}{0.5mm}
\setlength{\tabcolsep}{10pt}
\renewcommand{\arraystretch}{1.5}
\begin{table*}[ht!]
\begin{center}
\caption{Common Communication Technologies used in IoT-based Wearables}
\begin{tabular}{  |m{7em} | m{8em}| m{8em} | m{8em} | m{10em}|  }
  \hline
 Type of Technology & Range & Data Rates & Frequency Band (s)  &  Topology  \\
  \hline
  ZigBee & 10 to 100 meters & 250kbps & 2GHz & Star, ad hoc,
and mesh \\
  \hline
  WiFi & 10 to 100 meters & 6.75Gbps & 2.4GHz, 5GHz & Point Hub\\
  \hline
  Bluetooth & 10 to 100 meters & 2.1Mbps & 2.402GHz to 2.408GHz & Point to point, point to multi-point \\
  \hline
  LoRaWan & 15 to 20 Kilometers & 250bps to 5.5kbps & 169 MHz (Asia), 868MHz (Europe) 91MHz (North America) & Star \\
  \hline
\end{tabular}
\label{table:commT}
\end{center}
\end{table*}

\subsection{Safety-based wearables}
The wearables in the safety category are used to provide a safe environment for users. For example, mines can benefit from the use of safety-based wearable devices that monitor air quality to protect workers and reduce costs incurred by employers and the workers. Safety-based wearable devices are used in many applications to detect or prevent falls, especially in elderly people. Several studies in the literature have investigated and proposed numerous techniques and algorithms for fall detection. 

Using pretested templates for each type of fall and comparing angular velocities and angles between falls and normal activities, \cite{nyan2008wearable} propose a fall detection framework. In \cite{min2018detection}, deep learning and activity characteristics are used to develop a novel method for detecting human falls on furniture using scene analysis that adopts Region-based Convolutional Neural Network (R-CNN). A fall detection system incorporated tri-axial accelerometers, gyroscopes, and magnetometers to obtain fall parameters was proposed by \cite{pierleoni2015high}. Based on a reverse approach, the authors in \cite{ozcan2016autonomous} propose a novel autonomous fall detection system. The proposed system uses a camera attached to the subject rather than static sensors at fixed locations. Therefore, monitoring is not limited to areas where the sensors are located but includes all areas where the subject travels. \cite{ozcan2014automatic} presents a novel system that uses a smart camera to detect falls and classify activities around the waist. An EEG-based BCI prototype is proposed \cite{dhole2019novel} to detect whether an on-site worker is sleep-deprived or not in an elegant manner. Modified safety helmets with a discreetly placed signal acquisition device are worn by workers to ensure their safety. Table~\ref{table:SDPZTA} presents a summary of the categorized wearables and the relevant related works published in the literature.

%\subsection{Sports and Activity Recognition Applications}

%\subsection{Tracking and Localization Applications}

\section{Sensors in IoT-based Wearable}

Sensors are a crucial part of IoT-based wearable, they allow for the data collection of a certain action that can then be analyzed and measured. They are many different sensors, and each single sensor can have multiple different applications. 

\subsection{Sensors found in Sports}

Wearables that are used in sports to gain metrics and about an athlete's performance are categorized under Inertial Measurement Unit (IMU) sensors and help with the study of motions, kinematics \cite{sens1} The sensors that are categorized under IMUs are magnetometer, accelerometer, and gyroscopes. Magnetometer helps determine orientation of an object which in sports has many applications, the sensors themselves are small require low energy to power and can communicate over I2C or SPI \cite{bosch}. Accelerometers can measure movement in a given direction, they can be as precise to measure movement such as wrist tilt \cite{bosh1}. Gyroscopes measure orientation as well as angular velocity, it also has the ability to maintain orientation as well \cite{bosh2}. Combining the sensors into a single device can help measure and study kinematics in a sport setting. Examples of the sensors being used in a device can be found in ~\cite{golf} and \cite{basketball} where the sensors are used for their advantages in capturing certain ranges of motions. The data is then captured and analyzed by their respective data analysis programs. Sensors in IoT are very violate especially in Wearables, many sensors have more than one application, for example IMU sensors have applications in wearables outside of sports. For example, in \cite{3dimu} IMU sensors are used to create 3D maps, an application that is used outside of the scope of motion. 

\subsection{Sensors found in Healthcare }

When it comes to healthcare there is a long list of sensors that are found in wearables. Wearables in health care can be used to monitor or treat patients, which are powered by different types of sensors. Sensors already discussed \cite{bosch}\cite{bosh1}\cite{bosh2} also have use in healthcare since they deal with the study of motion with the human body \cite{sens1}. 

\setlength{\arrayrulewidth}{0.5mm}
\setlength{\tabcolsep}{10pt}
\renewcommand{\arraystretch}{1.5}
\begin{table*}[ht!]
\begin{center}
\caption{Sensors Found in Healthcare IOT Wearables}
\begin{tabular}{  |m{15em} | m{7em}| m{20em}|  }
  \hline
 Sensor & Product & Description   \\
  \hline
  Airflow Sensor\cite{airflow} & \cite{vent}& Helps correct a patients irregular breathing to supply enough oxygen to the body  \\
  \hline
  Pressure Sensor\cite{pressure} & \cite{vent}\cite{o2therapy}\cite{sleeptherapy}\cite{druginfusion} & Regulates and creates pressure in various applications \\
  \hline
  Oxygen Sensor\cite{sp02} & \cite{vent}\cite{pulseo2} & Measures a patients oxygen level as well as pulse rate\\
  \hline
  Electroencephalography(EEG)\cite{EEGSensor} Sensor & \cite{EEGprodcut} & Measures a patients brain activity\\
  \hline
  Glucose Sensor & \cite{GLProduct} & Measures a patients blood sugar level\\
  \hline
  Electrocardiogram(ECG) Sensor \cite{ECG} & \cite{ECG1}\cite{ECG2}\cite{ECG3} & Measures a patients heart rate\\
  \hline
  
\end{tabular}
\label{table:healthT}
\end{center}
\end{table*}

Looking at table~\ref{table:healthT} there are many different sensors found in different wearable devices for healthcare. Many sensors are present in many different devices, and sometimes are even combined with other sensors from other categories of wearables. Looking at the airflow sensor, it has a straightforward purpose within healthcare, to help a patient breathe. They can be found mostly in ventilators\cite{vent} as well as devices for oxygen therapy\cite{o2therapy}. A pressure sensor have an unexpectedly large impact on devices, they are used in ventilators, oxygen therapy, sleep therapy\cite{sleeptherapy}, and automation of drug infusion\cite{druginfusion}. Another sensor that is common throughout healthcare is oxygen sensors, which measure a patients oxygen level. The sensor can be found in devices that attach to a patient's finger \cite{pulseo2}. EEG sensors are also used in healthcare, they are used in research for the human brain to help gain a better understanding and can also be used to help diagnose any issues.
Glucose sensors mostly benefit patients with diabetes, the sensor has become non-invasive and implemented on IoT devices such as smartwatches that can automatically send alerts to let the patient know any imbalances are occurring. The ECG sensor might be the most know sensor in healthcare, it is present in every hospital and almost all smart watches/ring\cite{ECG1}\cite{ECG2}\cite{ECG3} feature an ECG sensor. The ECG measures heart rate and can collect information about a patient throughout the day to see if any irregularities are occurring. 

\subsection{Sensors found in Safety}
Wearables in the area of safety are also finding applications, the scenarios range from at home to work environment safety. Wearable devices are fitted with sensors that can check the environment, make sure certain equipment is being handled properly, and prevent and kind of injury. Taking a look at ~\cite{safteysensor} the authors used different sensors in a wearable device to determine harmful environmental conditions. The sensors used in their wearable are temperature, C02, UV, and CO ~\cite{ss1}\cite{ss2}\cite{ss3}\cite{ss4}. When it comes to safety most sensors can be implemented in wearables, for sports, health, and tracking and localization they can be added as preventative measures. For example, in sports, accelerometers can be used to detected impacts which can cause injuries, in healthcare ECG sensors can be used to detect abnormal heart rates, and in tracking sensors can be implemented to prevent someone from getting lost. 

\begin{figure}[h]
    \centering
    \includegraphics[width=.5\textwidth]{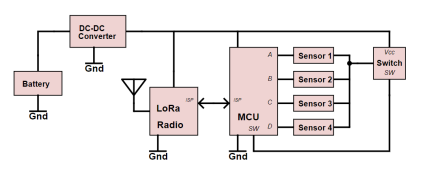}
    \caption{Example of Safety Wearable with sensors\cite{safteysensor}}
    \label{fig:safteysens}
\end{figure}

\subsection{Sensors found in Tracking/Localization}
Werables that are categorized in tracking and localization help with determining the environment conditions by utilizing sensors as well determining location \cite{local}. Tracking is a category with a wide area of applications, it can range from actual geo-locating to tracking a person's movements or specific body part. From a medical/sport standpoint sensor like accelerometers\cite{bosh1} can track movement to help determine the kind of motion, this also applies to magnetometer and gyroscopes. For determine location via satellite location sensors are used in wearables \cite{GPS}, this helps with everything between loss prevention to navigation. For more local mapping that does not need communication to satellite, like somewhere in a house or small community, accelerometers are used to measure distance between certain points to create a map. This comes into play in applications with smart homes, as well as mobile devices to help predict a user's habits.

\begin{figure}[h]
    \centering
    \includegraphics[width=.5\textwidth]{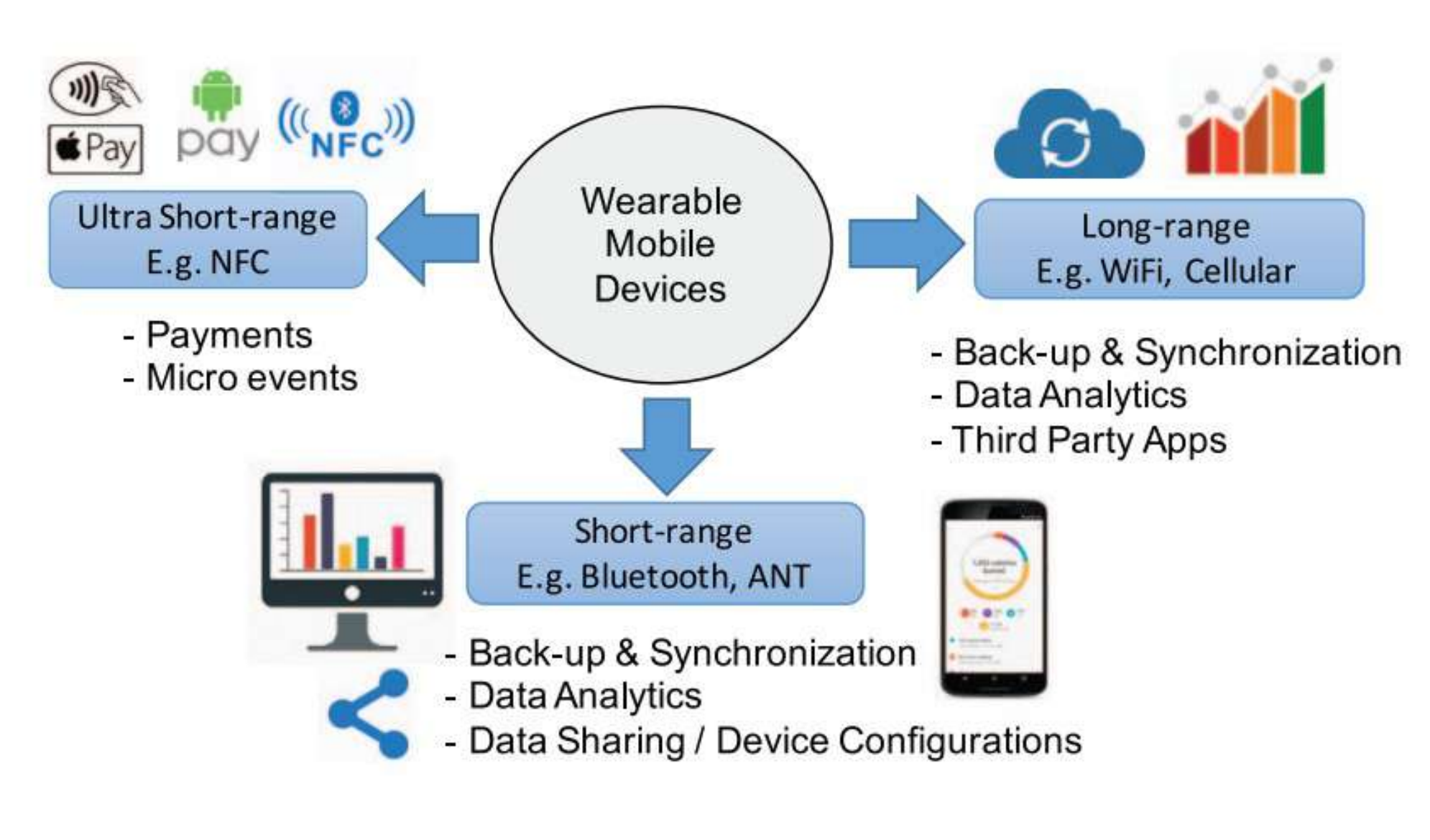}
    \caption{Communication Technologies for IoT-based Wearables \cite{seneviratne2017survey}}
    \label{fig:commTech}
\end{figure}

\section{Data Analysis and Communication technologies for IoT-based Wearables}

In this section, we will first introduce the various communication technologies that are widely adopted in IoT-based wearables. Then the various data analytic techniques in the context of wearables will be discussed.

\subsection{Communication Technologies}

Various wireless communication protocols support different communication ranges in wearables to offer a wide range of applications, as shown in Figure~\ref{fig:commTech}. Bluetooth, ZigBee and ANT belong to the short range class, Wi-Fi and cellular, LoRaWan belongs to the long range class and the ultra short range class consists of NFC.

\subsubsection{ZigBee}
ZigBee technology is built on the IEEE 802.15.4 standard. Resource-constrained environments and devices with limited power provide the perfect environment for this kind of technology to thrive. In spite of being standardized in 2003, it was originally conceived in 1998 \cite{zigbee}. ZigBee has a range of approximately ten to a hundred meters and consumes an insignificant amount of power. ZigBee uses the direct sequence spread spectrum (DSSS) technique and adopts the ad hoc, mesh and star topology. In ZigBee, there are three types of devices: the coordinator (ZC), the router (ZR), and the end-device (ZED) \cite{lounis2020attacks}.

\subsubsection{Wi-Fi}
A wireless communication standard based on IEEE802.11, wireless fidelity (Wi-Fi) is one of the most popular wireless technologies. In order to access radio channels, Wi-Fi uses the CSMA/CA protocol. With a range up to 10 meters and a high -power consumption, Wi-Fi is a highly efficient wireless network. There are four types of configurations available with this network: infrastructure, ad hoc, bridge, and repeater, which has a maximum data transfer rate of 6.75Gbps and 140MHz of channel capacity for a very high price \cite{henry2002wifi}.

\subsubsection{Bluetooth}
Wireless personal area networks (WPANs) with high security can be formed using Bluetooth, a proprietary open standard developed for mobile and fixed devices. Using a master-slave (client-server) model, Bluetooth is a packet-oriented protocol that is connection-oriented, packet-based. Bluetooth were limited by the initial specifications' high-power consumption, long communication set-up time, and high transmission latency (about 100 ms). Due to these limitations, Bluetooth Low Energy (BLE) was developed to overcome them. BLE is designed specifically for applications relating to health, sports, and fitness applications.

\subsubsection{LoRaWan}
IoT networks have become increasingly interested in LoRaWan (low-power wide-area networks) because it is suitable for LPWAN (low-power wide-area networks). LoRaWan is very suitable for wireless sensor networks due to its robustness and range. Frequency Shift Keying (FSK) or Chirp Spread Spectrum (CSS) are the main modes of communication used in LoRa radios. A node and a gateway are the two main types of devices in this technology. Gateways are connected to thousands of nodes at once, each sending and receiving information from the gateway in a LoRaWan setup.

In the wearable device domain, Bluetooth Classic (standardized but no longer maintained by IEEE 802.15.1), Bluetooth Smart and Wi-Fi have become the most common standards \cite{gravina2020wearable} used for connectivity to the Internet. A comparative analysis of the different adopted communication technologies used in IoT-based wearables in terms of range, data rates, frequency bands and topology is presented in Table~\ref{table:commT}.

\subsection{Data Analytics}
 Data analytics has a large application within wearables in IoT, since many new methods of machine learning are being developed, they can be used within these applications. In a system there is the device with its many sensors taking constant measurements it is the analytics that take the measurements and make them a readable output. IoT-based Wearables are classified in to four applications, health applications, sports applications, safety applications, tracking and localization applications. Wearables from each application can differ slightly in design and include the same sensors, it is how the data is analyzed that differs from each application. For example, for tracking and localization wearable can feature an accelerometers to track distance traveled but a wearable from healthcare application could be using an accelerometers to track the motion of a patient for physical therapy.

Machine learning can play an important role in data analysis for wearables. It can help learn a person's habits and find what is normal for the user. In \cite{damedical} it comes in the form of personalized medicine, where a machine learning algorithm can learn about a patient and then keep checks on the patient for irregularities outside of their normal. The same type of concept can be used in all applications of wearables. Another example of data analysis comes from \cite{damed2} where a wearable is being used to help diagnose early stages of Parkinson's disease. The sensors collect data which is then analyzed and compared to the early signs of the disease. 
\begin{figure}[h]
    \centering
    \includegraphics[width=.5\textwidth]{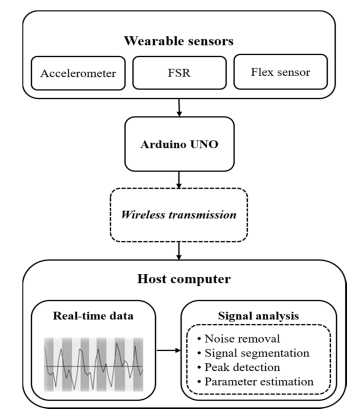}
    \caption{System used for collecting data and analyzing data from a sensor \cite{damed2}}
    \label{fig:damd}
\end{figure}

In many of the systems, the data analysis doesn't take place on the device itself. The computation needed cannot be met, there edge nodes and cloud -based systems are put in place to analyze the data. However, this doesn't come without challenges, many forms of data analytics require a large amount of data also many sensors are collecting data constantly throughout the entire day. Over time this accumulates into terabytes of data that can fill up storage drive quickly. Another challenge that must be addressed in privacy, the data that is being collected about a person's habits/lifestyle is very detailed and can be shared between different companies, making privacy an issue. In \cite{daprivacy} an approach is taken with data analytics to unlabeled the data to help privacy. By doing this they are not disclosing where the data is coming from and keeping to anonymous, the only competent that is receiving the labelled data is the algorithm doing the analysis. 

\begin{figure}[h]
    \centering
    \includegraphics[width=.5\textwidth]{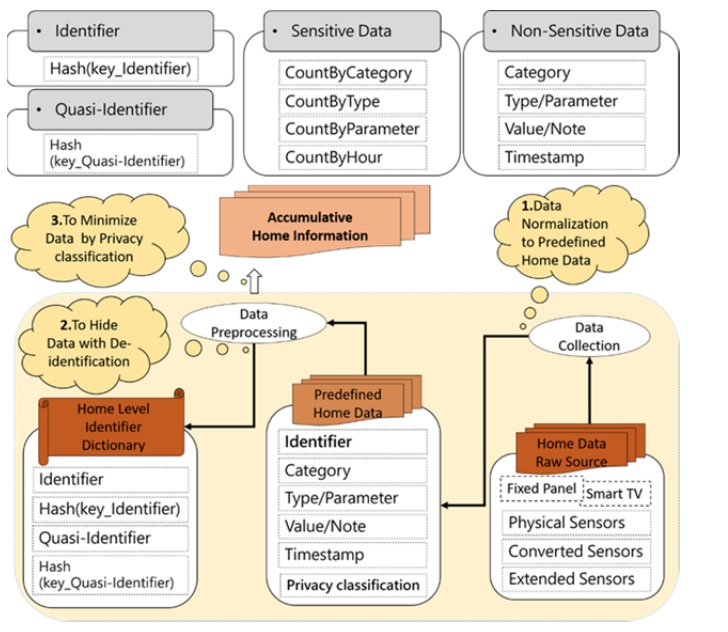}
    \caption{System used for privacy data processing \cite{daprivacy}}
    \label{fig:dapr}
\end{figure}

\section{Challenges and Future Research Directions}

There are several challenges in the domain of IoT-based wearables. To fully utilize the benefits of wearables devices in the community, these challenges need to be addressed urgently. 

\subsection{Battery-related Issues}

Since batteries have limited working times, wearable devices often have a short sustainable working time, which causes inconvenience to people's daily lives. Therefore, when designing wearables, special considerations need to be taken into account to minimize human interaction and ensure that the batteries last for many hours without needing to be replaced or recharged. For example, there are a variety of systems available, such as those that use low power consumption or techniques that harvest energy such as thermo-electric and piezo-electric that can be utilized in wearable domain.

\subsection{Communication-related Issues}

Currently, most wearable devices connect to the Internet primarily through proxy devices such as smartphones or personal computers. For example, most fitness tracking wearable applications still communicate with clouds through smartphones. This shows a gap in the adoption of direct communication with the internet, which hinders the development of some specific delay sensitive applications in the wearable domain. Potential reasons for this lack of support may be the absence of secure direct communication devices built into wearable operating systems or the slower development of third-party applications for wearables. However, the demand for wearable devices capable of direct communication with the internet is on the rise \cite{stats}. This is a potential area of research to explore the design of wearable devices with such capability.

\subsection{Trust-related Issues for Medical Use Cases}

There is lack of trust in the sensitive data produced by consumer wearables in patients monitoring applications. For example, data related to heart rate, pulse rate, and other health metrics is sensed by wearables using consumer hardware. Therefore, physicians are reluctant to use these data for critical diagnostics because they much rely on the accuracy of the hardware used in making the wearables as well as the accuracy of using it properly. A critical challenge would therefore be to improve health-related data accuracy cheaply.

\subsection{Privacy-related Issues}

It is possible that privacy breaches could occur as a result of an exchange of personal data between wearables and IoT hubs, including vital health signals, dosage, and location. It is typical for wearable IoT devices to operate in broadcast mode so that other network nodes can discover them easily. As a result, unauthorised users can intercept data in the form of eavesdropping attacks, which leads to privacy violation. A number of questions remain unanswered concerning the effective protection of users' privacy. This is an interesting research problem that is on-going not only in the wearable domain but in other security-related domains.

\section{Conclusion}

From fitness and sport to health monitoring, wearable devices are becoming increasingly popular. In this paper, we provided a comprehensive review of the most important research efforts from the literature in IoT-based wearables. We categorized the wearables according to their applicable applications. Additionally, the sensors, communication technologies and data analytic techniques adopted in the IoT-based wearables is investigated and presented by surveying multiple papers published in the literature. Additionally. the challenges as well as the future research directions in IoT-based wearables is presented. In terms of communication technologies, Bluetooth Classic, Bluetooth Smart and Wi-Fi are the most common standards adopted for wearables connectivity to the Internet.

\bibliographystyle{IEEEtran}
\bibliography{references}
\end{document}